\documentclass[aps,pre]{revtex4}
\pdfoutput=1

\usepackage[dvips]{graphicx}
\usepackage{amssymb,amsfonts,amsmath}
\usepackage{eucal,bm}
\usepackage{gensymb,subfigure}
\usepackage{color}
\usepackage{graphicx,graphicx,epsfig,color}

\newcommand{\qed}{\nobreak \ifvmode \relax \else
      \ifdim\lastskip<1.5em \hskip-\lastskip
      \hskip1.5em plus0em minus0.5em \fi \nobreak
      \vrule height0.75em width0.5em depth0.25em\fi}

\begin{document}

\title{Geometric Universality of Currents}

\author{Vladimir~Y.~Chernyak $^{a,b}$}
\author{Michael Chertkov $^{a,c}$}
\author{Nikolai~A.~Sinitsyn $^{a,c}$}

\affiliation{
$^a$Center for Nonlinear Studies and Theoretical Division, LANL, Los Alamos, NM 87545\\
$^b$Department of Chemistry, Wayne State University,
5101 Cass Ave,Detroit, MI 48202\\
$^c$New Mexico Consortium, Los Alamos, NM 87544}



\begin{abstract}
We discuss a non-equilibrium statistical system on a graph or network. Identical particles are injected, interact with each other, traverse, and
 leave the graph in a stochastic manner described in terms of Poisson rates, possibly dependent on time and instantaneous occupation numbers at the
 nodes of the graph. We show that under the assumption of  constancy of the relative rates, the system demonstrates a profound statistical symmetry,
resulting in geometric universality of the  statistics of the particle currents. This phenomenon applies broadly to many man-made and natural open stochastic
 systems, such as queuing of packages over the internet, transport of electrons and quasi-particles in mesoscopic systems, and chains of reactions in bio-chemical networks
. We illustrate the utility of our general approach using two enabling examples
 from the two latter disciplines.
\end{abstract}



\maketitle


Analyzing the statistics of currents generated in an open driven system that consists of multiple degrees of freedom is the holy grail of non-equilibrium statistical physics.
 The problem typically emerges in the context of multiple particles traversing a medium, subject to stochastic noise, and also interacting with each other. To study an elementary current
 means to count the number of particles passing through a respective line element of the medium during a fixed observation time.  Repeated a multiple number of times, these measurements of the
current show variable results due to the intrinsic stochastic nature of the underlying processes. To quantify these statistical variations one studies the Probability Distribution Function (PDF)
 of the current. It may also be useful and instructive to study simultaneously multiple currents, even the complete set of currents, related to all linear elements of the medium. When the
observation time is sufficiently long, in particular much
  longer than a typical time for a particle to traverse the system, one expects the PDF of the currents to obey a self-similar, so-called Large Deviation (LD) form, expressing the fact that,
 the longer the observation time is, the smaller the fluctuations in current density (total current accumulated over the observation time divided by the latter) are. The LD is the most general,
 yet not the only universal statement one can make about the PDF of currents. Thus, one popular formulation of the so-called Fluctuation Theorem (FT) suggests that under rather general conditions
the PDF of directional currents (which can be positive or negative, as counting the particles traversing a line element in different directions with different signs) also shows an additional profound
 symmetry,  e.g. relating to each other the statistics of large negative and large positive currents \cite{Evans93,95GC,98Kur,99LS,99Cro}. In this manuscript, we suggest a set of even more general
 (than FT) statements
  about the statistics of currents, observed in a variety of open many-body (i.e. consisting of multiple interacting particles) stochastic systems.

Examples of the systems, in which our consideration applies, are wide-spread in physics, biology, and operations research.  Fig.~\ref{fig:network_example} shows a graph of transitions between the nodes traversed by multiple jobs, cars, internet packets, or particles, which are processed/delayed at the nodes, e.g. competing for the servers' time, with the delay representing interactions for this example from Queuing Theory/Operations Research \cite{63Jac,76Kel,79Kel,01CY}. In statistical physics similar models are called zero-range processes \cite{70Spi,05ET}. Bio-chemical networks represent another area of application for the current (counting) statistics \cite{cox,dnelson-02pre,mfisher-motor-pnas,alon-elowitz,
low-copy1,low-copy2,low-copy3,english-00,orrit,gopich-03,english-06,xue-06,chaudhury-07,sinitsyn-09IET}.
The multiplicity of reactions for a given bio-chemical setting can be represented in terms of a graph of transitions between different bio-chemical states, e.g. as illustrated in Figs.~\ref{bio-pic}. Even when the graph is deduced reliably,  the transition (kinetic) rates often remain hidden, subject to inference/reconstruction from macroscopic measurements of counting statistics, which have recently extended the arsenal of experimental techniques available in the field. Yet another application domain which requires very precise evaluation of statistics of currents belongs to the field of mesoscopic electronics, e.g. electric circuits in systems made of quantum dots and tunnel junctions \cite{levitov,reznikov, qd-exp1,qd-exp2,turnstile,qd-exp1,qd-exp2,nazarov-03,gustavsson-06,sukhorukov-07Nat}, with an example illustrated in Fig.~\ref{QD-network}. Better understanding of the electron transport/current in such devices is instrumental for designing fast and reliable information
  processing systems for the future. These applications and examples will help us to illustrate the power of our main result, stated and explained in the next Section.

\section{Main Result: Geometric Universality of Currents}

\begin{figure}
\includegraphics[width=3in,page=1]{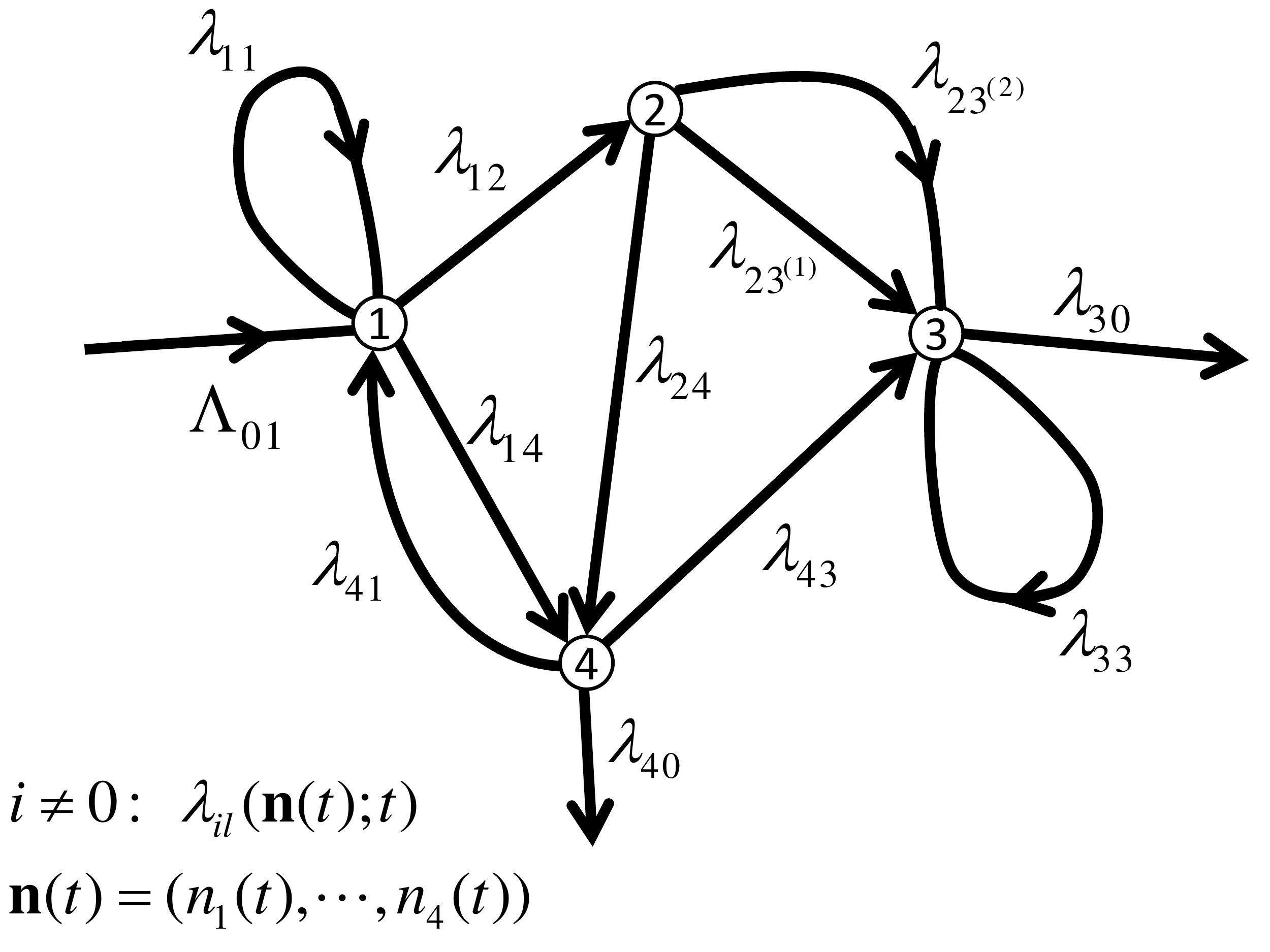}
 \caption{\label{fig:network_example}
 Example of an open network of interacting particles represented by a directed graph. The sample graph consists of four vortices/stations, labeled $1,2,3,4$, with label $0$ reserved for an external (reservoir) node.  Transitions between the nodes are shown as directed edges. Loops (self-loops, as $1\to 1$) are allowed. Each graph edge $(i,k)$ with $i\neq 0$ is equipped with a transition rate $\lambda_{ik}$, describing the rate of particle departure from site $i$ per current occupation number $n_i$ on the site. In this Letter we consider  the most general case of $\lambda_{ik}({\bm n}(t);t)$ being finite, and otherwise depending arbitrarily on time and the instantaneous global state ${\bm n}(t)$ of the system. $\lambda_{0i}$ stands for the time-independent rate of injection from the reservoir. We also allow two or more directed edges to connect the same vertex. In this context we mark the edges by distinct subindexes, like edges $23^1$ and $23^2$ in the Figure.}
 \end{figure}

In this manuscript we consider identical particles traversing the system, described by a connected open directed graph, ${\cal G}=({\cal G}_0,{\cal G}_1)$, for example shown in Fig.~\ref{fig:network_example}, where ${\cal G}_0$ and ${\cal G}_1$ denote the sets of the graph vertices and edges, respectively.  Particles enter the system from the exterior reservoir of infinite capacity, where $\lambda_{0i}$ is a time-independent Poisson particle injection rate at node $i$. Upon arrival at an entry node the particle advances by jumping between the nodes along the directed edges, each characterized by the transition rate, $\Lambda_{ik}={n}_i\lambda_{ik}(t;{\bm n}(t))$, with $\lambda_{ik}$ being the rate per particle that occupies node $i$. The internode per-particle rate, $\lambda_{ik}$, is finite and may depend on time explicitly, as well as implicitly via the dependence on the system state ${\bm n}=(n_i|i\in {\cal G}_0)$, with $n_i(t)$ being the number of particles at node $i$ at time
  $t$. We assume no intrinsic limitation on the number of particles residing at any node in the system, i.e. $n_i$ may be any nonnegative integer. A particle may leave the system along any of the outgoing links, with the corresponding departure rates $\Lambda_{i0}={n}_i\lambda_{i0}$, the per-node rate $\lambda_{i0}$ being also generally dependent on both ${\bm n}(t)$ and $t$. Such a stochastic system of many identical interacting particles is described via the
following Master Equation (ME) for the instantaneous distribution function of ${\bm n}$:
\begin{eqnarray}
&& \frac{\partial}{\partial t} P({\bm n};t)=\sum_{(0,i)\in{\cal G}_1} \lambda_{0i}\left(P({\bm n}_{-i};t)-P({\bm n};t)\right)\label{ME}\\
&& +\sum_{(i,j)\in{\cal G}_1}^{i\neq 0}\Biggl(
\Lambda_{ik}(t;{\bm n}_{+i})
P({\bm n}_{+i;-k};t)-\Lambda_{ij}(t;{\bm n})
P({\bm n};t)\Biggr),
\nonumber
\end{eqnarray}
where ${\bm n}_{+i}=(\cdots,n_i+1,\cdots)$, ${\bm n}_{-i}=(\cdots,n_i-1,\cdots)$ and ${\bm n}_{+i;-k}=(\cdots,n_i+1,\cdots,n_k-1,\cdots)$. We assume that the graph and the rates are chosen so
 that the solution of Eq.~(\ref{ME}) converges to a non-singular quasi-steady distribution of ${\bm n}$ in some finite time, $\tau$, which can be also viewed as the typical time a particle spends
in the system. To characterize the non-equilibrium features of this steady state one may also want to know the statistics of the vector ${\bm J}=(J_{ik}|(i,k)\in{\cal G}_1)$ of currents, with
 the edge-labeled components, $J_{ik}$, defined as the number of particles, that traverse through the edge $(i,k)$ during time $T$. For the observation time, $T$, that is long compared to the correlation time,
 $\tau$, the PDF of the currents obeys the following Large Deviation (LD) form: $-\ln({\cal P}({\bm J}|T))/T={\cal S}({\bm J}/T)+o(\tau/T)$, where ${\cal S}({\bm J}/T)$, often referred to as the LD function,
is a convex function of its multi-variant argument (in some physics and mathematical physics literature it may be also referred to as the Cr\'{a}mer function).

The main statement of this manuscript, which we call the {\bf Geometric Universality of Currents} (GUC), suggests that for $T\gg \tau$ and under the condition that the tested currents are not very large (so that the average ${\bm n}$ conditioned to the value of currents is non-singular and finite), {\it the LD function is invariant with respect to any (e.g. time and ${\bm n}(t)$ dependent) transformations that keep the graph intact and the branching ratios
\begin{eqnarray}
\forall (i,k)\in{\cal G}_{1}|\; i\neq 0:\quad \frac{\lambda_{ik}(t;{\bm n}(t))}{\sum_l^{(i,l)\in{\cal G}_1}\lambda_{il}(t;{\bm n}(t))}=\theta_{ik},
\label{rel_rates}
\end{eqnarray}
constant}, i.e. the transition probabilities $\theta_{ik}$ are independent of the current time $t$ and the current state ${\bm n}(t)$ of the network. In particular, the statement means that ${\cal S}({\bm J}/t)$ can be calculated by considering an auxiliary problem with $\lambda_{ik}(t;{\bm n})\to \theta_{ik}$, thus replacing the interacting system with its much simpler non-interacting surrogate. This reduction makes the evaluation of the LD function computationally easy. We label the universality, geometric, to emphasize its interpretation in terms of the statistics of open single-particle trajectories, and specifically of their geometries.
 Some geometric aspects of this analysis were already discussed in \cite{10CCGT} for a particular case of a queuing model with time-independent rates.

Our proposed {\bf GUC} also allows generalization to the case in which the injection rates dependend on time arbitrarily, for example in a quasi-periodic fashion, so that the time averaged
 rates $\bar{\lambda}_{0i}\equiv T^{-1}\int_0^Tdt \lambda_{0i}(t)$ are finite. Then, under the same conditions [universality and independence of time for all $\theta$'s in Eq.~(\ref{rel_rates}),
 and finiteness of the queues] as for the above statement of GUC, we find that {\it statistics of currents is universal with the LD function calculated, as if the injection rates were
 constant and equal to  $\bar{\lambda}_{0i}$}. We refer to this extension of GUC as {\bf the GUC for dynamic pumping}.

In what follows we will: (a) Discuss a set of examples from the literature, e.g. on queuing networks, networks of bio-chemical reactions, and electron transport over mesoscopic networks, in which many-body setting and associated statistics of currents are of interest. (b) Sketch a proof of the GUC. (c) Illustrate applications of GUC to bio-chemical networks and systems of quantum dots. (d) Conclude the manuscript by describing possible extensions and generalizations of GUC to other characteristics (such as the joint distribution functions of occupations and currents) and more general settings.

\section{Models}

We begin this Section by explaining a basic setting in the Operation Research (Queuing Networks) and then proceed with a discussion of our two enabling applications: one from the field of bio-chemical networks, and the other being related to electron currents in mesoscopic systems of quantum dots.

{\bf Queuing networks.} Introduced and studied in the work of Jackson \cite{63Jac} and Kelly \cite{76Kel}, this subject became classical in Operations Research (OR), see e.g. the textbooks \cite{79Kel,01CY}. Even though stated in somewhat different terms, the basic equation of the field, concerned with maintaining finite multiple queues in industrial conveyors, calling centers, transportation, and internet systems, is still the ME (\ref{ME}) with a specific choice of the rate dependence on the vector ${\bm n}$ of queues. In particular, the so-called Jackson network model assumes $\lambda_{ik}$ to be equal to a time-independent constant for $n_i\leq m_i$ and zero otherwise, with $m_i$ interpreted as the number of servers available at  node $i$. The particles (jobs, calls, cars, or packets) compete for the servers that may be busy serving other particles and thus interact, unless the $m_i$ are infinite at all nodes of the network. In the latter case the particles are ``free'',  i.e.
  they do not interact. The queuing network is called stable if a statistically steady solution of ME (\ref{ME}) exists, with all queues being finite. It was realized quite early in the history of this field \cite{76Kel} that if a network is stable, the stationary PDF of queues becomes factorized into the so-called product form, $P_{\mbox{st}}({\bm n})=\prod_i p_i(n_i)$,  where $p_i$ are the marginal distributions at the nodes. A similar decomposition has been also discovered by Spitzer \cite{70Spi} in the related mathematical physics context of interacting Markov processes. Statistics of currents in Jackson networks was studied much less in the classical OR literature. See e.g. \cite{79Kel,81WV}. Thus, the LD function of currents in the (static) Jackson networks was analyzed only recently \cite{10CCGT}. In this regards, this manuscript extends the approach and technique developed in \cite{10CCGT} to a more general setting,  in which transition rates in
the ME (\ref{ME})
 depend not only on the occupation number of the outgoing vertex, but also on the global state of the system,  characterized by ${\bm n}(t)$, and may also depend explicitly on time.

{\bf Biochemical Networks.} When the number of molecules involved in a bio-chemical reaction is not too large,  e.g. when it is of the order or less then $100$, the effects of noise on reaction kinetics of a moderate-size system (often called a mesoscopic system) are significant. For example {\it in vivo} reactions inside living bacteria \cite{hwa,low-copy1,low-copy2,low-copy3} with hidden kinetics (reaction rates are not known) is an exemplary mesoscopic system of this type. Effects of fluctuations in such systems are often analyzed {\it in vitro} using single-molecule experiments and methods of fluorescence correlation spectroscopy \cite{english-00,orrit,gopich-03,english-06,xue-06,chaudhury-07,sinitsyn-09IET}. These and other related techniques try to reconstruct the complex stochastic mechanisms of biochemical networks. In particular, the techniques are capable of uncovering new microscopic information on reaction kinetics, inaccessible via standard macroscopic (bulk) experi
 ments. This is achieved by measuring variances and autocorrelation functions of the reaction events.
 Examples of observations, which became available with the invention of single-molecule techniques, include the  discovery of new (not known before) internal states of enzymes \cite{english-06}. The opportunities provided by these techniques are exciting, however they are also accompanied  by significant challenges in interpreting the statistical measurements. Direct numerical simulations of even simple biochemical processes are prohibitively expensive and inconclusive for reconstructing full statistics of reaction events \cite{sinitsyn-09pnas}, thus emphasizing the importance of alternative analytical methods. This manuscript contributes to the task of developing analytical methods to describe delicate and important fluctuation effects in mesoscopic bio-chemical reactions. We specifically consider the following exemplary model.

\begin{figure}
\centerline{\includegraphics[width=3.2in]{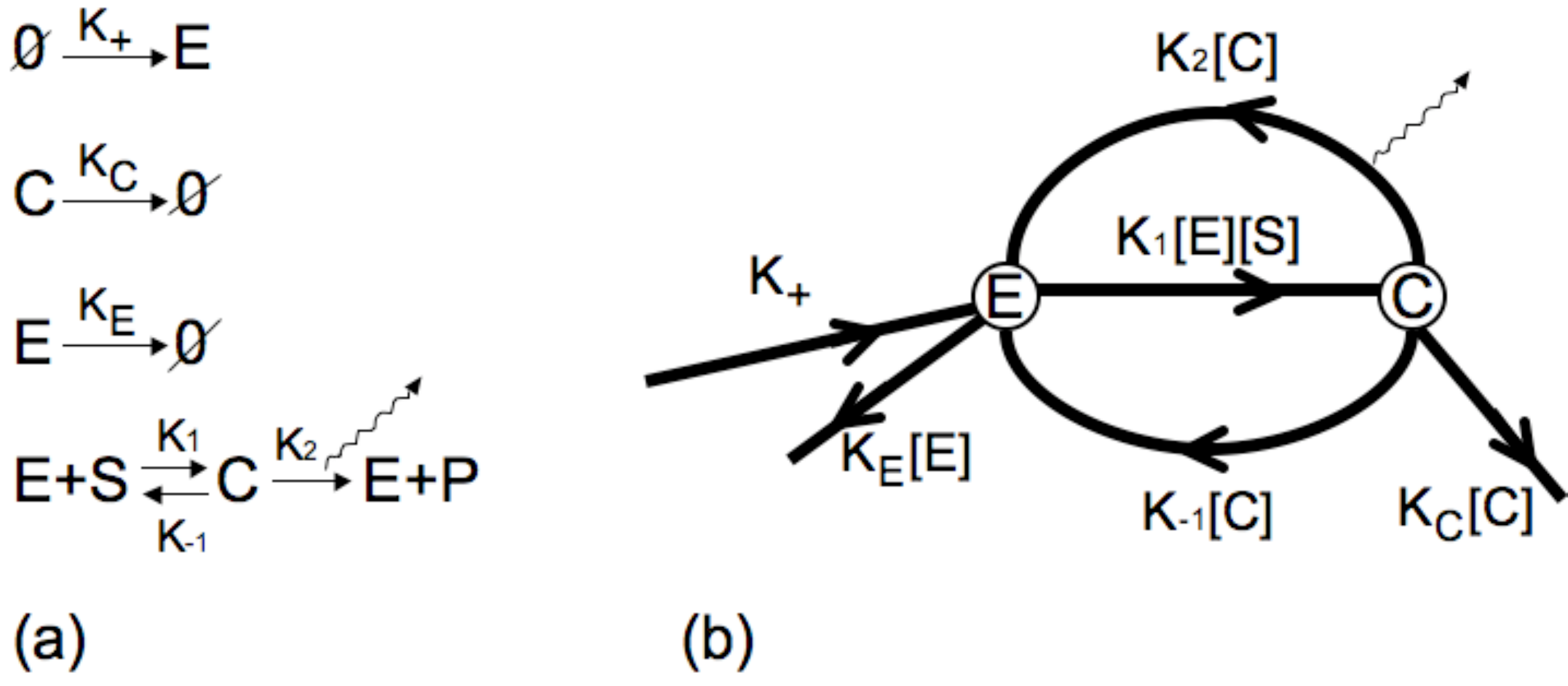}}
  \caption{ (a) Set of reactions that include a complex Michaelis-Menten process, as well as enzyme creation and degradation. (b) Equivalent Jackson network with kinetic rates which are proportional to the number of molecules, represented by the nodes. First and second nodes represent free enzymes and enzyme-substrate complexes, respectively.
  \label{bio-pic}
}
\end{figure}

{\bf Model 1: Noise in Enzymatic Reactions}: Consider the conversion of one type of molecules, called the substrate, into another type, called the product, via a complex  Michaelis-Menten (MM) enzymatic reaction \cite{MM}. This process involves interactions of a substrate molecule with another molecule, called the enzyme, that results in the creation of an enzyme-substrate complex. Let $S$, $P$, $E$, and $C$ denote the substrate, the product, the
enzyme, and the enzyme-substrate complex molecules, respectively. The  complex, $C$, is created from $E$ and $S$, followed by splitting into either $E+S$ or $E+P$. Assume the latter reaction is observable. This is possible, e.g. via attaching the so-called Green Fluorescent Protein (GFP) tag that fluoresces each time a product molecule is created \cite{english-06}. The light intensity is proportional to the number of product molecules. Measurements of the average intensity, its variance, and higher cumulants provides valuable information on the statistics of the number of transitions in the $C\rightarrow E+P$ sub-chain.
The enzyme molecules are also permanently generated and undergo degradation. 
To summarize,  the full set of elementary reactions that characterize the model is shown in Fig.~\ref{bio-pic}(a). It represents,
(1) creation of enzyme molecules, $0 \rightarrow E$, with rate $k_{+}$;
(2) degradation of the enzyme-substrate complex $C \rightarrow 0$ with rate $k_{C}[C]$;
(3) degradation of enzyme molecules, $E \rightarrow 0$, with rate $k_{E}[E]$;
(4) Conversion of $S$ into $P$ via a Michaelis Menten reaction, which involves three sub-processes:
(a) Creation of an enzyme-substrate complex $E+S \rightarrow C$ with rate  $k_{1}[E][S]$;
(b) Reverse reaction of a complex, splitting into free enzyme and substrate molecules, $C\rightarrow E+S$, with rate $k_{-1}[C]$;
(c) Irreversible splitting of a complex into free enzyme and a product molecules, $C\rightarrow E+P$, with rate $k_2[C]$.
Here $[\dots]$ represents a standard bio-chemical notation for abundance \cite{sinitsyn-09pnas}, i.e. the number of molecules of the given type in the system. We assume that the substrate molecules are supplied in macroscopic quantities, thus keeping the corresponding concentration $[S]={\rm const}$. Note that the described stochastic model of bio-chemical kinetics can be viewed in terms of stochastic transitions in the reaction graph shown in Fig.~\ref{bio-pic}(b), where the number of transitions through a given link represents the number of reaction events
during the observation time. Assuming all transitions to be Poisson, we observe that the set of bio-chemical reactions in Fig.~\ref{bio-pic}(a) can be viewed as an instance of a general stochastic network model with the property (\ref{rel_rates}). Specifically, the relations between the parameters of the bio-chemical models with fluctuating numbers of enzymes and the notation for the rates in our general network model (see e.g. Eq.~(\ref{ME})) are as follows: $\Lambda_{01}=k_+$, $\Lambda_{10}=k_{E}[E]$, $\Lambda_{02}=0$,
$\Lambda_{20}=k_{C}[C]$, $\Lambda_{12}=k_1[E][S]$,
$\Lambda_{21^1}=k_{-1}[C]$, $\Lambda_{21^2}=k_2[C]$.

\begin{figure}
\centerline{\includegraphics[width=2.82in]{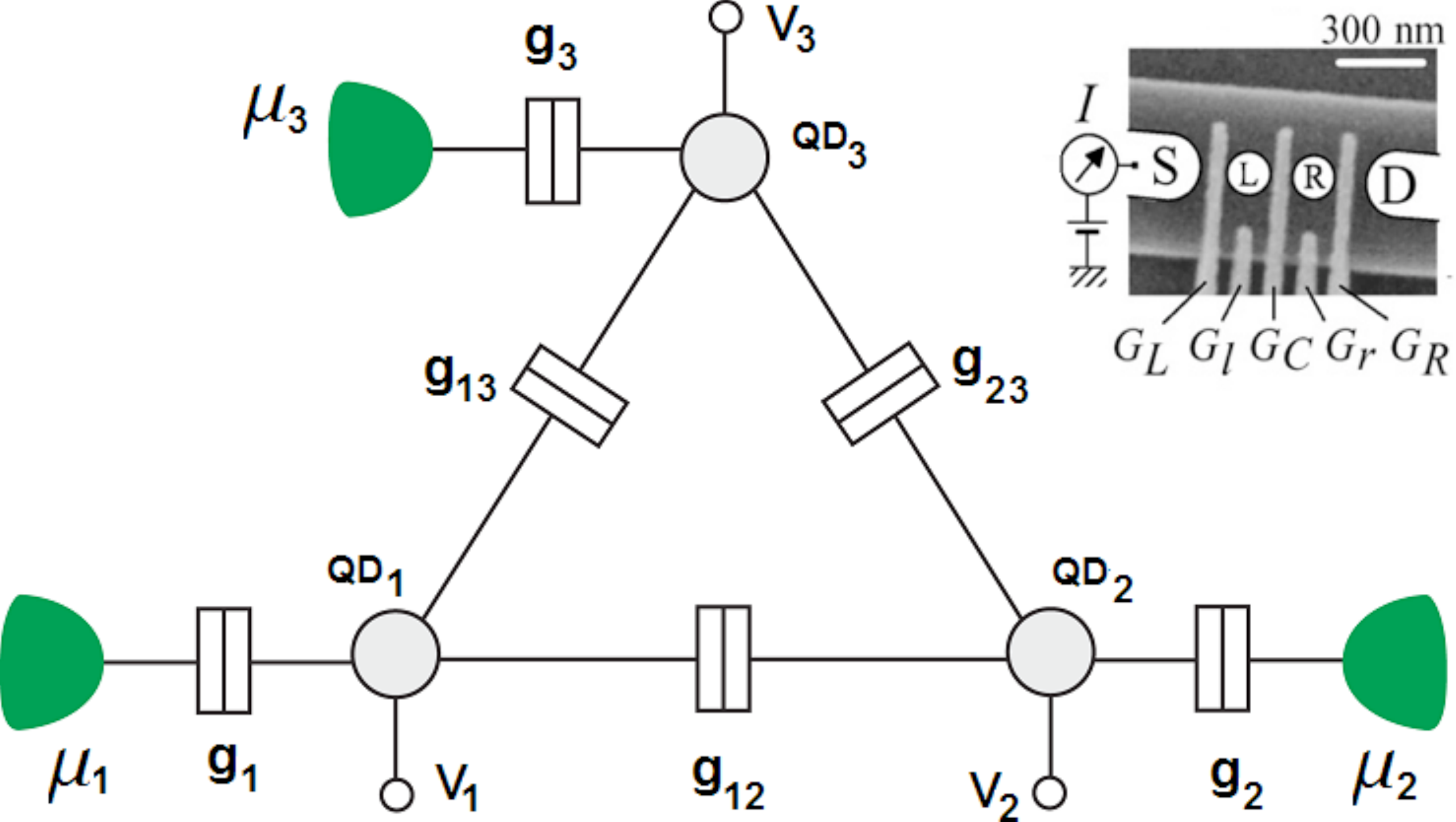}}
  \caption{ Electric circuit made of three quantum dots (QD), controlled by gate voltages, $V_i$, ($i=1,2,3$). Quantum dots are coupled to each other and to the leads via tunnel junctions with barriers, characterized by the parameters, $g_{ij}=g_{ji}$ and $g_i$, respectively. Leads (green) are characterized by their chemical potentials, $\mu_i$.
Iinset: Experimental realization of a double quantum dot from Fig. 7(a) in Ref.~\cite{qd-review}. Electron transport between source (S) and drain (D) proceeds through tunneling via two dots with potentials, tunable by gate voltages, $G_l$ and $G_r$. The barrier strengths at the tunnel junctions between two dots and between the dots and the leads S and D are controlled by additional gate voltages $G_L$, $G_R$ and $G_C$.\label{QD-network}
}
\end{figure}
\begin{figure}
\centerline{\includegraphics[width=3in,page=4]{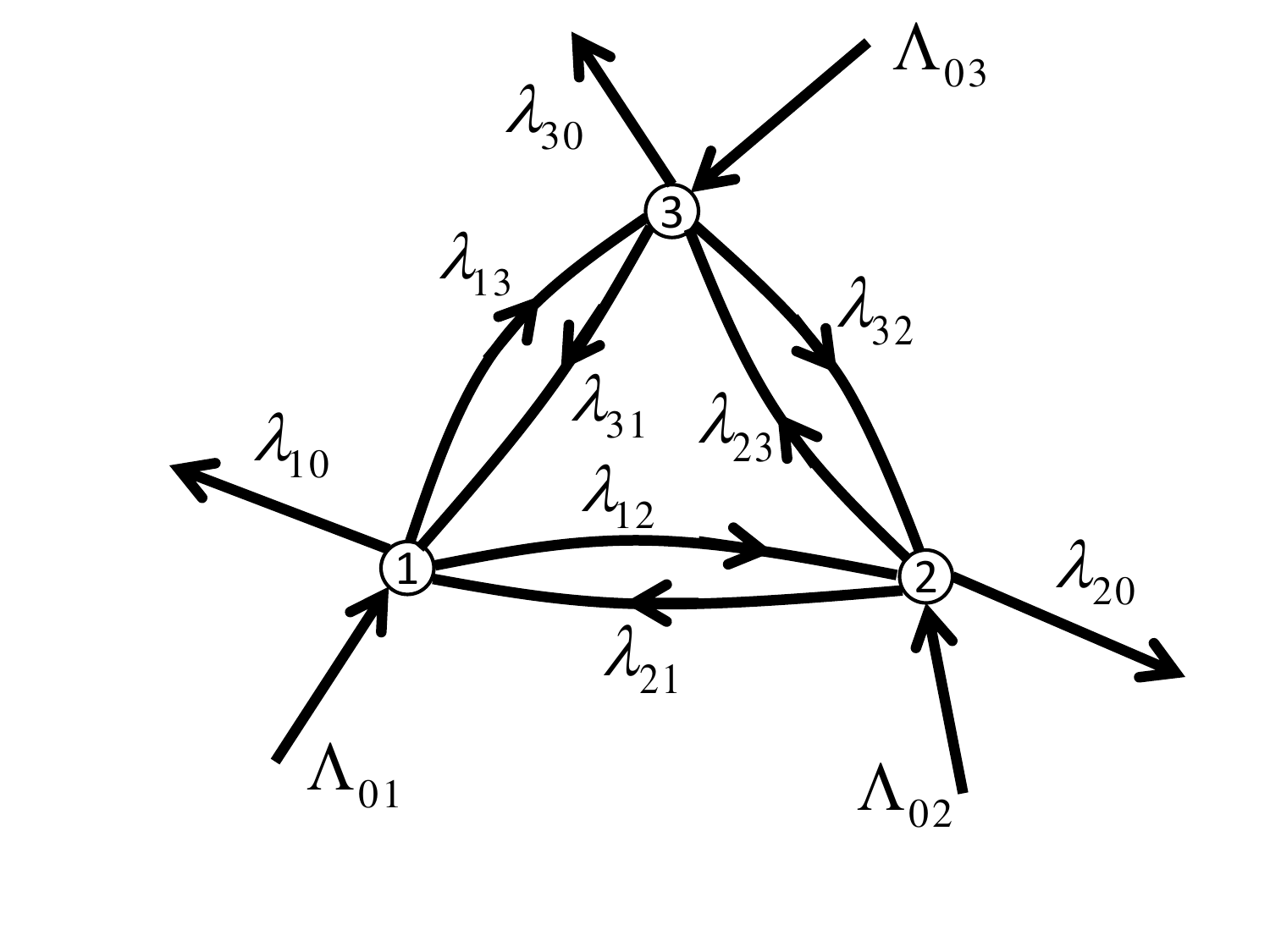}}
 \caption{\label{fig:DB-triangle}
 ``Queuing-style'' (oriented) graph representation for a system of three quantum dots shown in Fig.~\ref{QD-network}.  }
\end{figure}

{\bf Transport in Driven Mesoscopic Systems with Detailed Balance.}
Many mesoscopic systems in physics and chemistry can be modeled in terms of stochastic transitions that satisfy the Detailed Balance (DB) constraints with
 the participating particles entering and leaving the system via a reservoir, or a system of reservoirs. DB guarantees that for time-independent kinetic rates
 the system attains the equilibrium Boltzmann-Gibbs distribution. In a system with DB, an allowed transition from state $i$ to state $j$ is always accompanied
 by the reversed transition. A transition between two states $i$ and $j$ can be viewed in terms of a barrier, with the transition strength characterized by a
 symmetric characteristic $g_{ij}=g_{ji}$, independent of the system state ${\bm n}(t)$ (number of particles at a given node at a given time). Nodes are
 represented by the local trap potentials with  energies, $E_{i}$ and $E_{j}$, that control the asymmetry between the forward and backward transitions.
Energies may generally depend on the state (which represents
 multi-particle interactions) and time. It is convenient to adopt the Arrhenius parameterization of the DB per-particle kinetic rates transitions
 $\lambda_{ij}=\varepsilon_i g_{ij}$ and $\lambda_{ji}=\varepsilon_j g_{ij}$ for the $(i,j)$ and $(j,i)$ transitions, respectively, with $\varepsilon_i=e^{\beta E_i}$,
 and $\beta$ being the inverse temperature in  energy units. We also assume that all nodes of the open system are exchanging particles with the (external) reservoir.
 In fact, it may be convenient to introduce individual reservoirs for the nodes of the ``state'' graph, and characterize them in terms of their chemical potentials, $\mu_i$,
 and the barrier strengths, $g_i$. By analogy with the inter-node transitions we assume $g_i$ to be time- and state-independent, while the chemical potentials, $\mu_i$, may
 depend on time. Kinetic rates for transitions from node $i$ to its reservoir and vice versa are $\lambda_{i0} = \varepsilon_i g_i$ and $\Lambda_{0i} = \epsilon_i g_i
 $, respectively, with $\epsilon_{i}\equiv e^{\beta \mu_i}$.

For systems with DB, described above, the symmetry property (\ref{rel_rates}) is naturally built in, provided the barrier parameters  $g_{ij}$ and $g_i$ are kept time- and state-independent. Stated differently, the barrier parameters are single-particle, i.e. non-interacting, characteristics. Many-body interactions are represented by a (possibly complex and generally arbitrary) dependence of the node activation rates ${ \bm \varepsilon}=(\varepsilon_1, \varepsilon_2, \ldots, \varepsilon_N)$, on the system state ${\bm n}(t)$. Moreover, the non-equilibrium/driven nature of the system state originates from additional explicit time dependence of the activation rates ${ \bm \varepsilon}=(\varepsilon_1, \varepsilon_2, \ldots, \varepsilon_N)$ and ${\bm \epsilon}=(\epsilon_1,\epsilon_2,\ldots \epsilon_N)$ for the nodes and reservoirs, respectively. Periodic or quasi-periodic driving are two possible and representative examples. Measurements of the full counting statistics of currents
 in driven nanoscale devices, with otherwise DB conditions on kinetic rates, represent a feasible state-of-the-art in mesoscopic experiments in electronics \cite{mukamel-1,gustavsson-06, sukhorukov-07Nat} and soft matter \cite{gawedzki}.

{\bf  Model 2:} Transport in a network with DB is implemented in an electric circuit, such as in Fig.~\ref{QD-network}, that includes a number of semiconductor quantum dots populated by electrons which are allowed to hop through the tunnel junctions between the quantum dots and corresponding reservoirs. A  multi-reservoir queuing-style graph representation that corresponds to the electric circuit in Fig.~\ref{QD-network} is shown in  Fig.~\ref{fig:DB-triangle}. Electric circuits made of several quantum dots with tunable parameters have been implemented experimentally \cite{qd-exp1,qd-exp2,turnstile}. Such systems are used to create new devices capable of few-electrons information processing \cite{qd-exp1}, as well as showing useful features  of quantum dot ratchets \cite{qd-exp2}.

Our main focus for Model 2 is the possibility of observing the so-called {\it pump effect} that refers to efficient generation of currents induced by external periodic changes of the parameters, such as gate voltages that control the charging energies, $E_i$, in the quantum dots \cite{mukamel-2}.  The pump effect can be used to build a device, called an electronic turnstile \cite{turnstile}, that involves a circuit with quantum dots, driven by periodic gate voltages at zero bias between the leads. When the gate voltages are fixed, the electric current is not generated, however, under proper conditions, periodic driving generates currents, whose magnitudes can be controlled with high precision, thanks to the experimental ability to control the number of electrons pumped per cycle of the driving protocol \cite{astumian-quantized}. We consider such periodically driven electronic circuits in the regime of sufficiently high temperatures, so that the number of available states inside a quantum dot is large compared to a typical number of electrons in a dot. In this regime quantum coherence effects and exclusion interactions due to Pauli principle can be disregarded, so that electronic transitions through tunnel junctions are thermally activated and the corresponding rates can be described using the Arrhenius parametrization. Kinetic rates are controlled by application of gate voltages to quantum dots, as shown in Fig.~\ref{QD-network}. Note, that our forthcoming discussion of currents in the model accounts for many-body effects, originating, e.g., from Coulomb electrostatic interactions.

\section{Materials and Methods}

This Section has two parts.  We first give a sketch of the GUC proof and then illustrate the utility of our general result for two enabling applications, related to bio-chemical networks, and electron transport in quantum dots, respectively.

\subsection{Proof of GUC}

Our main result, GUC, can be rationalized in two different ways. The Lagrangian (Dynamic) approach is discussed below, while the complementary Eulerian (Static) consideration and the proof of GUC generalization to Dynamic Pumping are given in Section I of SI.

\subsubsection{Lagrangian (Dynamic) Approach}

\begin{figure}[t]
\includegraphics[width=1.5in,page=2]{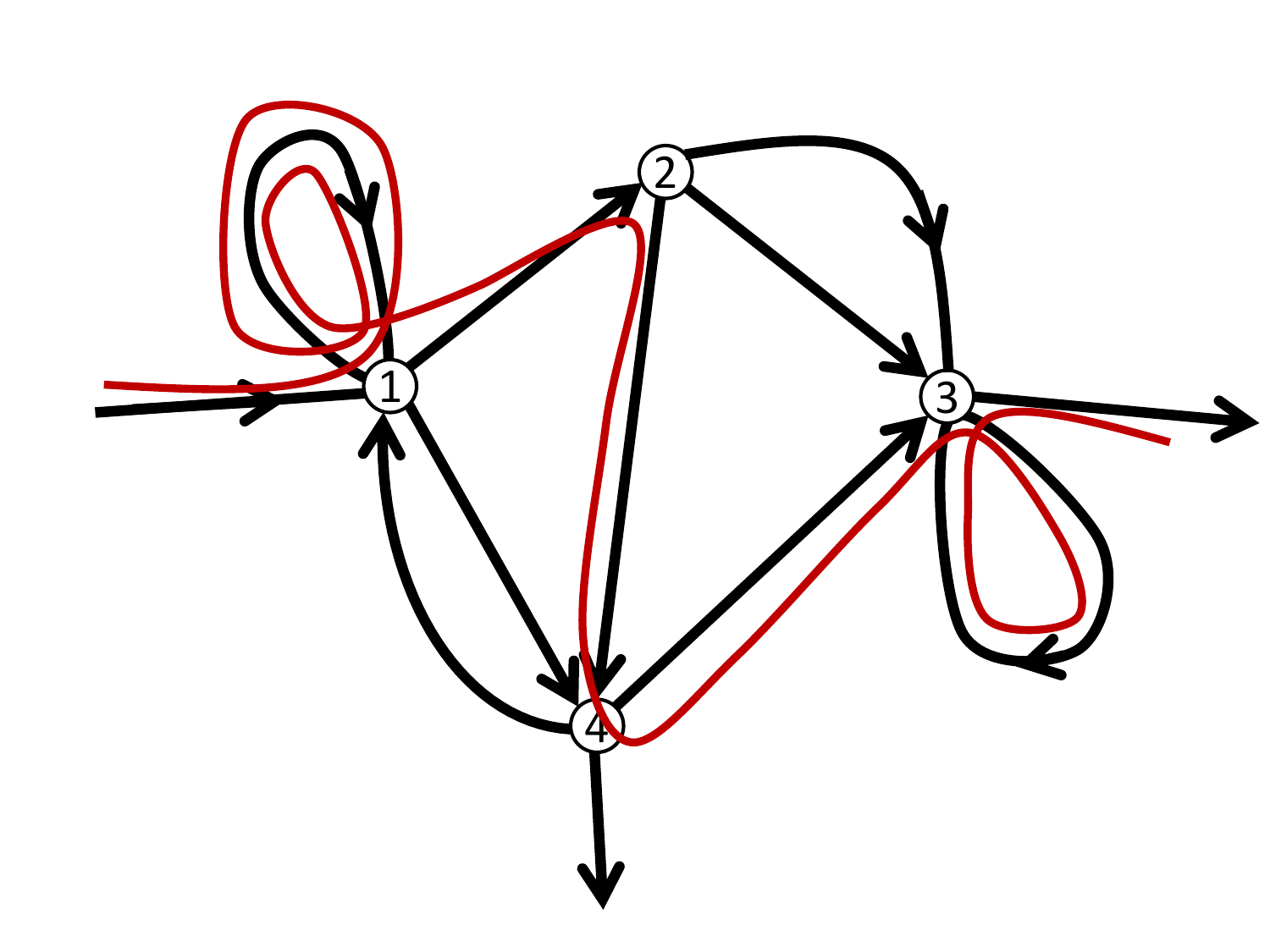}
\includegraphics[width=1.5in,page=3]{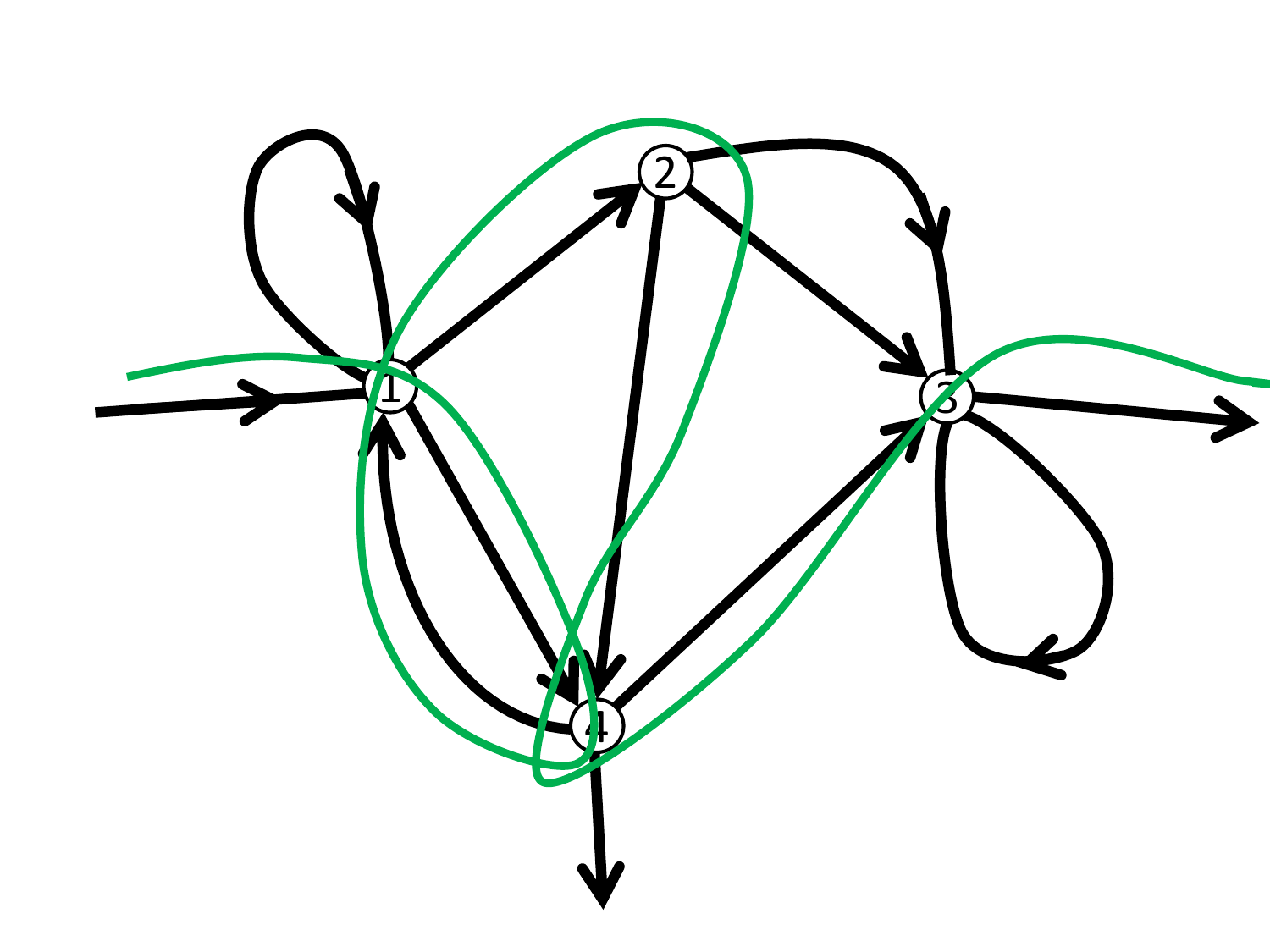}
\caption{\label{fig:paths} Two geometrically distinct paths for single-particle trajectories.
}
\end{figure}

Although in the interacting case the particle trajectories are strongly correlated, the main focus of our Lagrangian consideration will be on a single-particle trajectory,
 ${\bm x}=({\bm s},{\bm t})$, fully characterized by the following finite sequences, ${\bm s}=(s_{1},\ldots,s_{k})$ and ${\bm t}=(t_{0},t_{1}\ldots,t_{k})$, where $s_{l}$ and
 $t_{l}$, with $1\le l \le k$, stand for position (node $s_{l}\in {\cal G}_0$) and time stamp of the particle leaving the station $s_{l}$ for the next station $s_{l+1}$ along the path.
 We assume that $t_{0}$ and $t_{k}$ are the times when the particle enters and leaves the system, respectively; the temporal stamps are time-ordered: $0\leq t_{0}\leq\cdots\leq t_{k}\leq T$,
 where $T\gg\tau$ is a sufficiently long  time and $\tau$ is the correlation time (estimated as the typical time a particle spends inside the system); a particle can visit the
same station a multiple number of times; and the total number, $k$, of jumps can be an arbitrary
integer. It will be important for the forthcoming discussion to separate the geometrical, ${\bm s}=(s_{1},\cdots,s_{k})$, and temporal, ${\bm t}=(t_{0},\cdots,t_{k})$ characteristics
 of the trajectory. In particular, the total current generated by a single-particle trajectory, whose individual (per edge) component measures the number of times the trajectory passes
 through any given edge, depends on its geometrical component, ${\bm s}$, only, and the current generated by a stochastic realization of our system is the sum of the single-trajectory currents.
 Thus, to describe a single-particle trajectory, for example one of the two shown in Fig.~\ref{fig:paths}, we naturally pose the following question: what is the probability for a particle
observed entering and leaving the network over sufficiently long time horizon $[0;T]$ to choose this particular path, ${\bm s}$? This probability should depend on the factors $\theta_{s_{l}s_{l+1}}$
 only and, therefore, it shows no correlations
  with other trajectories. The information contained in this probability distribution, ${\cal P}({\bm s})$ is geometric and independent of the temporal details ${\bm t}$ of the trajectory.
The assumed invariance (\ref{rel_rates}) of our general model, with respect to any changes in the system keeping the transition (branching) probabilities $\theta_{il}$  constant, guarantees
that the probability of a path, ${\cal P}({\bm s})=\prod_{l=1}^{k} \theta_{s_{l}s_{l+1}}$ (with $s_{l+1}=0$), is universal/invariant as well. The probability of a temporal realization of a
many-particle trajectory shows strong correlations between single-particle trajectories and can be explicitly expressed in terms of $\kappa_{s_{l}}(t;{\bm n}(t))\equiv \sum_{i\in {\cal G}_{0}}^{(s_{l},i)\in{\cal G}_{1}}\lambda_{s_{l}i}$,
 however, the generated currents are not sensitive to this complex temporal structure.

We now consider the probability of observing a given vector of the current for the observation time, $T$ (the component of the vector associated with an edge counts the number of particles that have crossed the edge in time $T$.)
\begin{eqnarray}
{\cal P}({\bm J}|T)&=&\sum_{n=1}^{\infty}\sum_{{\bm s}^{1}\ldots{\bm s}^{n}}{\cal P}({\bm s})\prod_{(i,k)\in {\cal G}_1}\delta(J_{ik}-\chi_{ik}({\bm s})) \nonumber \\ &\times& {\cal P}_{0}(s_{0}^{1},\ldots,s_{0}^{n}) ,
\label{PJ}
\end{eqnarray}
where ${\cal P}_{0}(s_{0}^{1},\ldots,s_{0}^{n})$ is the probability of $n$ particles being injected into the network at the corresponding nodes and $\chi_{ik}({\bm s})$ simply counts the number of times the particles traversing the path ${\bm s}$ went through the edge $(i,k)$. Obviously, Eq.~(\ref{PJ}) is evaluated as an expectation over all possible single-particle paths. [Eq.~(\ref{PJ}) is an asymptotic expression, valid for $T\gg\tau$ only, because it completely ignores the trajectories, trapped in the system for  periods longer than $T$]. Therefore, one concludes that asymptotically ${\cal P}({\bm J}|T)$ is a single-particle object which is also universal/invariant with respect to Eq.~(\ref{rel_rates}) in view of its linear relation to ${\cal P}({\bm s})$ and the previously discussed universality of the latter. \qed

\subsection{Applications of GUC}

Here we discuss two applications of GUC to Model 1 and Model 2. In the context of these two models GUC can be viewed as a no-go restriction on the ability to interpret experimental data on the statistics of reaction events (Model 1), or to control electrical currents in networks with DB (Model 2). These restrictions follow from the fact that current statistics actually depends only on the time-independent relative probabilities, $\theta_{ij}$ defined in Eq.~(\ref{rel_rates}) as ratios of transition rates, rather than the complete transition rates themselves.

{\bf Model 1.} Suppose that one performs measurements of the number of product molecules converted from the substrate by enzyme molecules via a set of enzymatic reactions. Assume that this experiment can be performed in a single-molecule setting and repeated to determine not only the average rate of the product creation, but also to measure and quantify mesoscopic fluctuations of the currents. One can attempt to use the measured data to extract information on the reaction
 mechanisms, e.g. to determine the kinetic rates for elementary reactions, involved in the enzymatic reaction.

A version of the general GUC relevant for this model is: {\it For steady (time-independent) rates of enzymes' creation and degradation, the counting statistics of the enzymatic reaction events does not provide full quantitative information on all kinetic rates of elementary reactions involved in the process}. The number of independent parameters that cannot be determined is, at least, equal to the number of different states of the enzyme and the enzyme-substrate complex. This restriction follows from the observation that if we multiply all outgoing rates from any node of the graph of transformations by a constant factor, the full counting statistics of transitions through any link in the network will not change. Therefore, for each network node there is a set  of parameters that cannot be extracted from the counting statistics. This general result is illustrated in Section B of SI using an example of the open-system Michaelis-Menten (MM) mod
 el of Fig.~\ref{bio-pic}. The main {\it no-go} (prohibitive) conclusion of the open-system MM analysis is that the MM curve for the average current (along the upper $C\to E$ link from Fig.~\ref{bio-pic}), presented as a function of $[S]$, depends only on two independent quantities, namely $k_2k_{+}/k_{-}$ and the MM constant $K_{\rm MM}=(k_2+k_{-1})/k_1$. Measurements of the Fano factor (defined as the ratio of the second-to-first cummulants of the current) provide additional information on the ratio, $k_2/k_{-}$, thus allowing the value of $k_+$ to be extracted from a combination of the two measurements. However, the other two rates of the complex,  $k_1$ and $k_{-1}$, cannot be determined even if $k_{-}$ is known independently from some complimentary experiment.

{\bf Model 2}: For a system with DB, it is convenient to consider undirected links that can be traversed in both directions, so that the current associated with an original (undirected) link is the difference between particles moving in the main (positive) and reverse (negative) directions. These currents will be referred to as  bi-directional currents. Obviously, statistics of bi-directional currents are obtained from $\Delta({\bm q})$ by setting  $q_{ij}=1/q_{ji}$ in the argument of the generating function. Then, according to GUC for dynamic pumping, the statistics for the vector of bi-directional currents, built of the components associated with undirected edges (we will still use ${\bm q}$ for the argument of the generating function, not to complicate the notation), is characterized by
\begin{eqnarray}
\Delta({\bm q }) =\sum_{i} g_i\bar{\epsilon}_{i}(\zeta_{j}/q_{i}-1),
\label{bra-CGF1}
\end{eqnarray}
where $\bar{\bm \epsilon}=({\bar \epsilon}_i=T^{-1} \int_0^{T}dt \epsilon_i(t)|i\in {\cal G}_0)$, and the set of ${\bm \zeta}=(\zeta_i|i\in{\cal G}_0)$ (defined at all graph vertices) satisfies
\begin{eqnarray}
\label{eigenvalue-bra-coherent} \forall k:\quad\sum_{i}g_{ik}(q_{ik}\zeta_{i}-\zeta_{k})-g_{k}\zeta_{k}=-g_{k}q_k.
\end{eqnarray}
Eq.~(\ref{bra-CGF1}) shows that in spite of the many-body interaction and explicit time dependence of the parameters ${\bm \varepsilon}$ and ${\bm \epsilon}$, the statistics of currents appears to be the same as in the auxiliary time-independent and non-interacting system with $\bar{\bm \epsilon}$ expressing the effective constant chemical potentials of the reservoirs. We arrive at the following general (as applied to any open DB system driven externally in a periodic or quasi-periodic fashion)
{\bf No-pumping Fluctuation Theorem} (NPFT) relating currents in the original and reduced systems:
{\it If the barrier characteristics $g_{ij}$ of the transitions, and $g_i$ of the reservoirs remain time-independent, while vectors ${\bm \varepsilon}$ and ${\bm \epsilon}$ of the activation rates are driven arbitrarily in time (periodically or quasi-periodically) and ${\bm \varepsilon}$ generally depends on the local node populations, then, provided the observation time is long compared to all correlation times in the system, the counting statistics of currents is indistinguishable from its counterpart in a reference non-interacting, ${\bm \varepsilon}={\bm 1}$, system that stays in its steady state (not driven) with $\bar{\bm \epsilon}$ being for the vector of the reservoir activation rates.} This general statement contains {\it no-pumping} in its name, since it prohibits generation of currents that are different from the steady state currents under the aforementioned conditions for periodic pumping. In particular, NPFT guarantees that if the vector $\bar{\bm \epsilon}$ of
 the reservoir activation rates has all the same entries, then all odd current cumulants remain zero, i.e. the current statistics are indistinguishable from their counterparts for the currents in thermodynamic equilibrium, even if the driving protocol explicitly breaks  time-reversal symmetry.

Extended discussions of the relation between NPFT and the previously known FT, as well as further illustrations of how NPFT can be used to build an efficient electronic turnstile \cite{turnstile}, are given in Sections II and III of SI, respectively.

\section{Conclusions and Path Forward}

In this manuscript we have formulated and demonstrated a strong symmetry, referred to as {\it Geometric Universality of Currents}. GUC imposes restrictions on stochastic of currents generated over long times in the systems of interacting particles in open networks. We have shown that the GUC rests on the geometrical nature of currents, and is limited to large but not extreme values of the currents, so that almost all particles spend only a finite time within the network. We have observed that GUC extends only to the LD function (the leading, exponential term in the PDF) while the pre-exponential factor is not universal.

 GUC imposes strong restrictions on the amount of information one can extract by measuring the statistics of currents. From this viewpoint, studying the conditions under which GUC can be violated and extending GUC to more
general settings can be expected to have increasingly important role in the future. This may take place via finding the dead ends and opening new venues for interpretations of experimental measurements in terms of the underlying stochastic mechanisms and microscopic parameters, as well as identifying efficient strategies for design and optimization of nanoscale electronic and biochemical devices. Therefore, we would like to conclude with a brief discussion of possible future generalizations.

The most general setting, considered in this manuscript still has three basic restrictions: the observed currents are not too large, the number of particles occupying a node is unrestricted, and particles are identical, of only one kind. It has been recently shown \cite{10CCGT} in the context of Jackson networks that even when the network is stable (i.e. when particles do not accumulate in the steady state), generation of atypically large currents is accompanied by accumulation of particles in the system. However, even in this regime of extreme currents, the LD form of the current distribution holds, ${\cal P}({\bm J}|T)\sim \exp(-T{\cal S}({\bm J}/T))$, and the LD function ${\cal S}({\bm j})$ remains well-defined. The most probable way to reach these atypically large values of ${\bm j}$ is associated with accumulation of particles at some, so-called saturated, nodes. In this regime of extreme currents a finite fraction of particles will not leave the system during the observa
 tion time (in spite of the fact that the latter is much larger than the correlation time $\tau$), thus violating one of the key assumptions used to derive GUC. However, and as shown in \cite{10CCGT}, the saturated nodes start acting as additional Poisson sources (reservoirs), thus leading to survival of GUC in a partial form -- with respect to variations in rates at the non-saturated nodes. To study this effect of the partial breakdown/survival of GUC under extreme currents is an interesting task for the future.

Introducing limitations on the number of particles that can occupy the same node, as well as analyzing several types of particles that interact with each other differently at different nodes (the two regimes are known in the queuing literature as one of the finite waiting room, and one characterized by graph-inhomogeneous priorities, respectively), makes the analysis of currents much more complicated. Addressing these issues in the context of GUC should become yet another challenge for the future. Finally, let us mention that the GUC analysis may also be extended evaluating LD functions of currents in various interesting non-equilibrium cases quantitatively,  e.g. in the spirit of \cite{07TCCP,09CCMT,10CCGT}.

\begin{acknowledgments}

The work at LANL was carried out under the auspices of the National Nuclear Security Administration of the U.S. Department of Energy at Los Alamos National Laboratory under Contract No. DE-AC52-06NA25396. This material is also based upon work supported in part by the National Science Foundation under CHE-0808910 at Wayne State U, and under EMT-0829945 and ECCS-0925618 at NMC.

\end{acknowledgments}
\newpage

\appendix

This supplementary material presents further details on the various results discussed in the main body of the paper, following the plan described hereafter.

Section \ref{sec:GUC} achieves two goals. First it describes in Section \ref{subsec:GUC_Eulerian} a complementary Eulerian (static) proof of the main result of the paper, Geometric Universality of Currents (GUC), and second it extends the Eulerian approach to prove {\it GUC for dynamic pumping}, which is done in Section \ref{subsec:GUC_dynamic}. Section \ref{sec:MM} describes in detail the application of GUC to the Michaelis-Menten process. We continue our discussion of the No-Pumping Fluctuation Theorem (NPFT) in Section \ref{sec:NPFT}, first commenting in Section \ref{subsec:NPFTasFT} on the relation between NPFT and the other, previously known, Fluctuation Theorems (FT), followed by outlining the possible application of NPFT to turnstiles in Section \ref{subsec:turnstile}.

\section{Further Discussions of Geometric Universality of Currents (GUC)}
\label{sec:GUC}

\subsection{Proof of Geometric Universality of Currents (GUC): Eulerian Approach}
\label{subsec:GUC_Eulerian}

Geometric universality of currents can be also demonstrated using the operator approach. The latter is implemented to study the generating function,
\begin{eqnarray}
\label{define-Z} Z({\bm q};T)=\sum_{{\bm J}}{\cal P}({\bm J}|T)
\prod_{(i,k)\in {\cal G}_{1}}q_{ik}^{J_{ik}},
\end{eqnarray}
depending on the multi-variant argument ${\bm q}=(q_{ik}|(i,k)\in{\cal G}_{1})$. We will focus on the analysis of the $T\gg\tau$ asymptote, where $Z({\bm q};T)\sim e^{T\Delta({\bm q})}$. Since ${\cal S}({\bm j})$, with ${\bm j}={\bm J}/T$, is related to $\Delta({\bm q})$ by the standard Legendre transform
\begin{eqnarray}
&& {\cal S}({\bm j})=\sum_{(i,k)\in{\cal G}_1}j_{ik}\ln q_{ik}^*-\Delta({\bm q}^*),
\label{Legendre}\\
&& \forall (i,k)\in{\cal G}_1:\quad j_{ik}=q_{ik}^* \partial_{q_{ik}}\Delta({\bm q}^*),
\nonumber
\end{eqnarray}
the universality of the LD function, ${\cal S}({\bm j})$, under the considered transformations is equivalent to the universality of $\Delta({\bm q})$.

Within the respective operator, so-called Doi-Peliti,  formalism \cite{76Doi,85Pel} (see also \cite{07Zei,10CCGT}) the Master Equation [see Eq.~(1) of the main body] becomes, $\partial P({\bm n};t)/\partial t=\hat{H}(t)P({\bm n};t)$, where the time-dependent operator $\hat{H}(t)$ is
\begin{eqnarray}
\label{M-operator} \hat{H}(t)&=&\sum_{(0,i)\in {\cal G}_{1}}\lambda_{0i}(\hat{a}_{i}^{\dagger}-1)+\sum_{(i,j)\in{\cal G}_{1}}^{i,k\ne 0}\theta_{ij}\hat{a}_{k}^{\dagger}\hat{b}_{i}(t) \nonumber \\ &+& \sum_{(i,0)\in {\cal G}_{1}}\theta_{i0}\hat{b}_{i}(t)-\sum_{k\in{\cal G}_{0}}^{k\ne 0}\hat{a}_{k}^{\dagger}\hat{b}_{k}(t),
\end{eqnarray}
with $\hat{a}_{i}^{\dagger}P({\bm n})\equiv P({\bm n}_{-i})$ and
\begin{eqnarray}
\label{define-a-b} \hat{b}_{k}(t)P({\bm n})\equiv (n_{j}+1)\kappa_{k}(t,{\bm n}_{+k})P({\bm n}_{+k}).
\end{eqnarray}
In deriving the last term in Eq.~(\ref{M-operator}) we used $\sum_{k}^{(i,k)\in{\cal G}_{1}}\theta_{ik}=1$. Note that $\hat{H}(t)$ has a very simple form, being expressed via the standard ``creation'' operators $\hat{a}_{k}^{\dagger}$, which are local and time-independent, and the non-standard ``annihilation'' operators $\hat{b}_{k}(t)$, which are non-local and depend on time explicitly. The complexity of the system stochastic dynamics is hidden in the time-dependent, nonlocal, and nonlinear nature of the ``annihilation'' operators.
 As we will see, the function, $\Delta({\bm q})$, is not sensitive to the specific form of $\hat{b}_{k}(t)$, which provides the Euler (operator) view of the GUC.

Extending the standard operator approach developed for generating functions in \cite{10CCGT} to the most general case, where the rates depend on the occupation numbers and time arbitrarily, we derive
\begin{eqnarray}
\label{Z-operator-form} Z({\bm q};T)=\sum_{{\bm n}}\hat{U}_{{\bm q}}P_{0}({\bm n}), \; \hat{U}_{{\bm q}}(T)=\mbox{Texp}\left(\int_{0}^{T}dt\hat{H}_{{\bm q}}(t)\right).
\end{eqnarray}
Here, $P_{0}({\bm n})$ is the initial distribution, $\mbox{Texp}$ stands for the time-ordered exponential, and the twisted master operator $\hat{H}_{{\bm q}}(t)$ is
\begin{eqnarray}
\label{M-operator-twisted} \hat{H}_{{\bm q}}(t)&=&\sum_{(0,i)\in {\cal G}_{1}}\lambda_{0i}(q_{0i}\hat{a}_{i}^{\dagger}-1)+\sum_{i\in{\cal G}_{0}}^{i\ne 0}\hat{c}_{i}^{\dagger}({\bm q})\hat{b}_{i}(t) \nonumber \\ \hat{c}_{i}^{\dagger}({\bm q})  &\equiv& \sum_{k\ne 0}^{(i,k)\in{\cal G}_{1}}q_{ik}\theta_{ik}\hat{a}_{k}^{\dagger}-\hat{a}_{i}^{\dagger}+q_{i0}\theta_{i0}.
\end{eqnarray}
In the long-time limit the spectral decomposition of Eq.~(\ref{Z-operator-form}) is dominated by the eigenvalue of the time-ordered exponential operator with the highest absolute value. Solving the eigenvalue problem $\hat{U}_{{\bm q}}|\Psi_{{\bm q}}\rangle=z_{{\bm q}}|\Psi_{{\bm q}}\rangle$ in the most general case does not appear feasible, thus reflecting the complexity of the most general stochastic dynamics of interacting particles. However, the non-trivial simplicity in the system that leads to the universal property of the generated currents shows itself when we study the eigenvalue problem for the bra-, rather than ket-, eigenstate: $\langle\Psi_{{\bm q}}|\hat{U}_{{\bm q}}=\langle\Psi_{{\bm q}}|z_{{\bm q}}$ . The bra-eigenvalue problem for the highest absolute value can be treated explicitly by observing that for given ${\bm q}$ the family of operators $\hat{H}_{{\bm q}}(t)$, parameterized by time, all share the same bra-eigenstate with the same eigenvalue, i.e., $\langle \Psi_{{\bm q}}|\hat{H}_{{\bm q}}(t)=\langle\Psi_{{\bm q}}|\omega_{{\bm q}}$, so that $z_{{\bm q}}=e^{\omega_{{\bm q}}T}$, which yields $\Delta({\bm q})=\omega_{{\bm q}}$.

The bra-eigenstate is represented by a coherent state $\langle{\bm\zeta}|\equiv \langle{\bm 0}|\prod_{j\in{\cal G}_{0}}e^{\zeta_{j}\hat{a}_{j}}$, with $\langle{\bm 0}|$ denoting the state with no particles. The coherent states are parameterized by ${\bm\zeta}=(\zeta_{j}|j\in {\cal G}_{0})$ and satisfy the following important property: $\langle{\bm\zeta}|\hat{a}_{j}^{\dagger}=\langle{\bm\zeta}|\zeta_{j}$. Due to the latter property a coherent state $\langle{\bm\zeta}|$ is an eigenstate of the first term on the rhs of Eq.~(\ref{M-operator-twisted}). Therefore, to ensure that $\langle{\bm\zeta}|$ is an eigenstate of $\hat{H}_{{\bm q}}(t)$ it is sufficient to verify that $\langle{\bm\zeta}|\hat{c}_{i}^{\dagger}=0$ for all $i$. Consequently, this condition results in a system of linear equations
\begin{eqnarray}
\label{linear-eq-zeta}
&& \sum_{k \ne 0}\theta_{ik}(q_{ik}\zeta_{k}-\zeta_{i})+\theta_{i0}(q_{i0}-\zeta_i)=0, \\
&& \omega_{{\bm q}}=\sum_{i}\lambda_{0i}(q_{0i}\zeta_{i}-1), \nonumber
\end{eqnarray}
whose solution fully identifies $\Delta({\bm q})=\omega_{{\bm q}}$. Finally, the GUC follows from the fact that (\ref{linear-eq-zeta}) depends only on the time-independent parameters, $\theta_{ik}$ and $\lambda_{0i}$. \qed

\subsection{Proof of GUC for Dynamic Pumping}
\label{subsec:GUC_dynamic}

It appears that the Eulerian derivation of the main GUC presented above allows for a straightforward and very simple generalization proving {\bf GUC for dynamic pumping}.  Indeed, the first equation in Eqs.~(\ref{linear-eq-zeta}) does not depend on the pumping rates $\lambda_{0i}$ at all. Thus $\langle\Psi_{{\bm q}}|=\langle{\bm\zeta}|$, with ${\bm\zeta}$ obtained by solving (\ref{linear-eq-zeta}), will also be independent of the pumping rates. Therefore, $\langle\Psi_{{\bm q}}|=\langle{\bm\zeta}|$ forms a bra-eigenstate for the time-ordered exponential, defined in (\ref{Z-operator-form})
\begin{eqnarray}
\langle\Psi_{{\bm q}}|\hat{U}_{{\bm q}}(T)=\langle\Psi_{{\bm q}}|\exp(\int_{0}^{T}dt\sum_{i}\lambda_{0i}(t)(q_{0i}\zeta_{i}-1).
\label{GUC_dyn_pumping}
\end{eqnarray}
This results in a universal expression, $\Delta({\bm q})=\bar{\omega}_{\bm q}=\sum_{i}\bar{\lambda}_{0i}(q_{0i}\zeta_{i}-1)$, for the general case of time-dependent pumping rates. Applying the Legendre transform (\ref{Legendre}) to $\Delta({\bm q})$, one arrives at the GUC for dynamic pumping.\qed

\section{Counting Statistics in the Michaelis-Menten Process}
\label{sec:MM}

As stated in the main body, the counting statistics of the enzymatic reaction events does not provide full quantitative information on all kinetic rates of elementary reactions involved in the process. Here we illustrate the general principle, using the model shown in Fig.~2 of the main text. We are interested in the counting statistics of transitions through the link $(2,1)^2$ only, thus setting all the generating factors  in Eq.~(\ref{linear-eq-zeta}), except $q_{21^1}$, to one. To simplify the notations, in this paragraph we will be using $q$ instead of $q_{21^2}$. Our focus is on the analysis of $\Delta(q)=k_{+}(\zeta_1-1)$, where $\zeta_1$ satisfies the following equation, which is a version of (\ref{linear-eq-zeta}),
\begin{eqnarray}
&& k_1[S] (\zeta_2-\zeta_1)-k_{E}\zeta_1=-k_{E},\nonumber\\
&& k_{-1}(\zeta_1-\zeta_2)+k_2(q\zeta_1-\zeta_2)-k_{C}\zeta_2=-k_{C}.
\label{sol-mm1}
\end{eqnarray}
This results in the following expression for the leading exponent in the generating function,
\begin{equation}
\Delta(q)=\frac{(q-1)k_1[S]k_2k_{+}}{k_1[S](k_2(1-q)-k_C)+k_{E}(k_2+k_{C}+k_{-1})}.
\label{cmg-mm-sol}
\end{equation}
This expression shows the dependence on the relative probabilities, $k_C/k_2$, $k_{-1}/k_2$ and $k_E/k_1[S]$ only, and it does not depend on the rates explicitly. Obviously, a similar statement can be made regarding the LD function of the current, defined according to the Legendre transform (\ref{Legendre}). To avoid bulky expressions we are not showing ${\cal S}(j)$ explicitly, but rather present the corresponding expressions for the average current and its variance (these are the two major characteristics currently available experimentally, while measuring the higher cumulants still represents a significant experimental challenge as it requires higher accuracy)
\begin{eqnarray}
\langle j\rangle & = & \left.q\partial_q\Delta(q)\right|_{q=1}\label{av-mm-sol}\\
&\approx & \frac{k_2[S](k_{+}/k_{-})}{[S]+K_{{\rm MM}}}, \quad K_{MM}=\frac{k_2+k_{-1}}{k_1},\nonumber
\\
\frac{1}{T}\langle (j-\langle j\rangle)^2\rangle &= &
\left(\partial_q q\partial_q\Delta(q) -(q\partial_q\Delta(q))^2\right)|_{q=1} \label{var-mm-sol} \\ & \approx &
\frac{2[S]^2k_2^2k_{+}}{k_{-}^2([S]+K_{{\rm MM}})^2}.
\nonumber
\end{eqnarray}
In deriving Eq.~(\ref{av-mm-sol}) we assumed that the enzyme and its complex degrade with the same rate $k_E=k_C\equiv k_{-}$, and considered the limit of slow degradation, $k_{-} \ll k_{-1},k_{2},k_1[S]$. The combination of parameters, denoted in Eqs.~(\ref{av-mm-sol},\ref{var-mm-sol}) by $K_{{\rm MM}}$, is called the Michaelis-Menten constant. Since the average number of enzymes is given by the ratio $k_{+}/k_{-}$,
the average current $\langle j\rangle$ in Eq.~(\ref{cmg-mm-sol}) is given by the famous Michaelis-Menten law for the average enzyme concentration. The effect of the enzyme number fluctuations on the variance of the product creation rate is, however, dramatic. It has been shown previously \cite{sinitsyn-09pnas} that the noise in an MM reaction with a fixed number of enzymes is suppressed, so that the so-called Fano factor $f \equiv T\langle (j-\langle j\rangle)^2\rangle/\langle j\rangle$ is smaller than unity, thus showing sub-Poisson statistics. In our case the Fano factor is given by
\begin{equation}
f= 1+\frac{2k_2[S]}{k_{-}([S]+K_{{\rm MM}})},
\label{noise-mm}
\end{equation}
i.e. the underlying statistics is super-Poissonian, with the Fano factor being much larger than one, as $k_2/k_{-}\ll 1$. From the Michaelis-Menten curve for the average current, presented as a function of $[S]$, one can measure only two independent quantities, namely $k_2k_{+}/k_{-}$ and the MM constant $K_{\rm MM}$.
Measurements of the Fano factor provide additional information on the ratio $k_2/k_{-}$, thus allowing the value of $k_+$ to be extracted from a combination of the two measurements. However, the other rates $k_1$ and $k_{-1}$ cannot be determined, even if $k_{-}$ is known independently from some complimentary experiment. It is instructive to compare these {\it no-go} results with the calculations of the Fano factor for the class of enzymatic mechanisms with a fixed number of enzyme molecules, considered in \cite{sinitsyn-09IET}, where it was shown that if the number of enzyme molecules is fixed, measurements of the Fano factor are sufficient to distinguish among different enzymatic mechanisms, even if on average they produce identical MM curves. Moreover, according to \cite{sinitsyn-09IET}, it was possible to determine the values of all the kinetic rates for the transitions among the metastable states within these complex reactions. Our result shows that for a wider class of
 reactions, such measurements of the reaction event statistics can be insufficient to fix the complete set of kinetic rates. Hence other statistical characteristics that are also sensitive to the temporal, rather than only geometric properties of stochastic trajectories (e.g. representing statistics of the time intervals between successive events), would be required.

\section{Further Discussions of No-Pumping Fluctuation Theorem (NPFT)}
\label{sec:NPFT}

\subsection{Relation between NPFT and previously known FT}
\label{subsec:NPFTasFT}

NPFT contains {\it fluctuation theorem} in its name because of its close relation to other fluctuation theorems formulated in the past for systems with DB. Recently, the statistical theory of fluctuations in mesoscopic systems has experienced considerable progress due to the discovery of exact results called fluctuation theorems \cite{Evans93,95GC,98Kur,99LS,kurchan1,falkovich} and nonequilibrium work relations \cite{bochkov-81,jarzynski-97prl,99Cro,05CCJ}, however, unlike NPFT, the previously found exact results describe or follow from the relations among standard thermodynamic characteristics such as entropy, work, or information. The NPFT represents a fluctuation theorem of a new type. It is related to but different from the recently discussed pumping-restriction theorems and related results stated for the average currents \cite{jarzynski-08prl,sinitsyn-08prl,netocny-10jcp,horowitz-09} in closed networks. The latter pumping-restriction theorems, however, cannot be generalized to current
  fluctuations, neither can they be derived as a special case of the solution of Eqs.~(17-18) of the main part of this paper.

Other, previously known, fluctuation theorems can be derived from the general solution of Eq.~(17) from the main body. If one starts with Eq.~(18) of the main body, divides it by $q_k$,
multiplies by $\epsilon_k=e^{\mu_k}$, and then sums over $k$, one finds that the rhs of the resulting expression coincides with Eq.~(17) of the main body. If we additionally assume that $q_i=e^{\alpha_i}$ and $q_{ij}=e^{ (\mu_j-\mu_k)-(\alpha_j-\alpha_k)}$, where $\alpha_i$, $i=1,\ldots N$ are arbitrary parameters, then the lhs will be identical to zero, since the expression under the $\sum_{ij}$ is odd with respect to a permutation of the indices. Considering the definition of $\Delta({\bm q})$, the result is equivalent to the following fluctuation theorem for the vector of the total number ${\bm J}$ of particles, transferred through the links
\begin{equation}
\langle e^{ \sum_i \alpha_i J_i + \sum_{i,j} [(\mu_i-\mu_j)-(\alpha_i-\alpha_j)]J_{ij} } \rangle = 1.
\label{ft-big}
\end{equation}
 (\ref{ft-big}) is a Jackson network generalization of the standard fluctuation theorem called the
Jarzynski equality \cite{jarzynski-97prl,99Cro,05CCJ}, $\langle e^{ -W  } \rangle = 1$, where $W$ is the total work performed by external  forces over the system. The Jarzynski equality is obtained from Eq.~(\ref{ft-big}) by setting $\alpha_i= \mu_i$, and noticing that $W \equiv \sum_i\mu_i J_i$.

\subsection{Application of NPFT to Electronic Turnstiles}
\label{subsec:turnstile}

The NPFT leads to important restrictions on the strategies for building an efficient electronic turnstile \cite{turnstile}, defined as a circuit of quantum dots driven by periodic gate voltages at zero bias between the leads. At fixed gate voltages, this system does not sustain an electric current on average. However, periodic driving may induce currents, whose magnitude can be controlled with high precision since it is possible to pump a specified number of electrons per cycle of the driving  protocol \cite{astumian-quantized}. Our solution describes such periodically driven electronic circuits in the regime of sufficiently large temperatures, when the number of available states inside quantum dots is large compared to the typical number of electrons. In this regime, quantum coherence effects and exclusion interactions due to Pauli principle can be disregarded, and electronic transitions through the tunnel junctions, separating the quantum dots from the leads and from each other,
are thermally activated. Kinetic rates are controlled by applying voltages to the leads or gate voltages to the quantum dots, as shown in Fig. 3. Many-body electron interactions inside the same quantum dots, e.g. correspondent to the Coulomb electrostatic interaction, are allowed. Our solution predicts that in the thermally dominated regime one cannot induce a directed current among the leads or tunnel junctions connecting quantum dots,
no matter how gate voltages are modulated. Moreover, currents, integrated over the driving protocol period, will show counting statistics that are indistinguishable from that observed in the thermodynamic equilibrium.

\bibliographystyle{pnas2009}
\bibliography{queuing-meso}


\end{document}